\documentclass[letterpaper, 10 pt, conference]{ieeeconf}  

\IEEEoverridecommandlockouts                              

\usepackage{graphicx}  
\usepackage{qcircuit}
\usepackage{amsmath}
\title{\LARGE \bf
The quantum version of the shifted power method and its application in quadratic binary optimization}

\author{Ammar Daskin$^{1}$ 
\thanks{$^{1}$Ammar Daskin is with the Department of Computer Engineering, Istanbul Medeniyet University,Uskudar, Istanbul, Turkey,
        {\tt\small ammar.daskin@medeniyet.edu.tr}}
}
\newcommand{\ket}[1]{\ensuremath{\left|#1\right\rangle}} 

\renewcommand{\bf}[1]{\ensuremath{\mathbf{#1}}}
\begin{document}

\maketitle
\thispagestyle{empty}
\pagestyle{empty}

\begin{abstract}

In this paper,  we present  a direct quantum adaptation of the classical shifted power method. 
The method is very similar to the iterative phase estimation algorithm; however, it does not require any initial estimate of an eigenvector and as in the classical case its convergence and the required number of iterations are directly related to the eigengap.
If the amount of the gap is  in the order of $1/poly(n)$, then the algorithm can converge to the dominant eigenvalue in $O(poly(n))$ time.    
The method can be potentially used for solving any eigenvalue related problem and finding minimum/maximum of a quantum state in lieu of Grover's search algorithm. 
In addition, if the solution space of an optimization problem with $n$ parameters is encoded as the eigenspace of an $2^n$ dimensional unitary operator in $O(poly(n))$ time and the eigengap is not too small, then the solution for such a problem can be found in $O(poly(n))$. 
As an example, using the quantum gates, we show how to generate the solution space of the quadratic unconstrained binary optimization as the eigenvectors of a diagonal unitary matrix and find the solution for the problem.
\end{abstract}

\section{Introduction}
On quantum computers, the quantum phase estimation algorithm \cite{nielsen2002quantum} can be used to compute an eigenpair of a given unitary matrix. 
The algorithm is the main part of the integer factoring\cite{shor1994algorithms} and HHL\cite{harrow2009quantum} algorithms and quantum chemistry calculations \cite{kassal2011simulating,brown2010using}.
In addition to adiabatic quantum computing \cite{farhi2000quantum,farhi2001quantum} as a global optimization method  and approximation algorithms such as \cite{farhi2014quantum};  the introduction of HHL algorithm for solving linear systems of equations has paved the way for more research in iterative methods. These methods so far include the quantum least square \cite{wossnig2018quantum}, the quantum versions of the gradient descent \cite{rebentrost2016quantum,kerenidis2017quantum}, and conjugate gradient method \cite{shao2018quantum} proposed for optimization problems.
There is also a considerable research interest for quantum semidefinite programming which can also be used for discrete optimization problems. Quantum speed-ups for semidefinite programming are reported in \cite{brandao2017quantum,van2017quantum} by forming  the solution as a quantum state. Ref.\cite{van2018convex} has considered the query complexity of a general convex optimization given by oracles and  provided the conditions to reduce the complexity on quantum computers. 
Recently, Shao \cite{shao2018quantum} has used the quantum version of the power iteration based on HHL algorithm  as a preliminary technique for the quantum Arnoldi algorithm. 

In this paper, without using HHL algorithm,  we sketch the steps of the shifted power method based on just measurement and Hadamard gates. 
In the following section, after explaining the classical shifted power method, we describe its quantum version. Then, we analyze its complexity, discuss how it can be used in the eigenvalaue problems, and present numerical simulations.
In Sec.III, we discuss possible applications. In particular we present a quantum circuit to map a discrete optimization problem such as quadratic binary optimization and quadratic assignment problems to a diagonal unitary matrix and use the quantum power method to solve these problems.
 
\section{Shifted power method}
The most basic numerical eigenvalue algorithms on classical computers are the power method and its variants such as the shifted power method \cite{saad2011numerical}.
The shifted power method is a well-known simple classical algorithm used to find the dominant eigenvalue and the eigenvector of a matrix. 
For a complex matrix $U$ of dimension $2^n$ with eigenvalues $\lambda_1, \lambda_2, \dots, \lambda_{2^n}$,  the shifted power method is described by the following algorithmic steps:
\begin{enumerate}
	\item An initial non-zero eigenvector $\bf{v_0}$ is chosen.
	\item For $k = 1\dots maxIterate$, the following computation is iterated until the convergence:
		\begin{equation}
		\bf{v_{k}}= \frac{(I-U)\bf{v_{k-1}}}{\alpha_k} 
		\text{, with\ }
		\alpha_k = \left\lVert{(I-U)\bf{v_{k-1}}}\right\rVert.  
		\end{equation}
\end{enumerate}
Here,  as $\alpha_k$ converges to the dominant eigenvalue of $(I-U)$, $\bf{v_{k}}$ converges to its associated eigenvector. 
If we assume that $|1-\lambda_1|\geq |1-\lambda_2| \geq \dots \geq |1-\lambda_{2^n}|$, then $\alpha_k\approx |1-\lambda_1|$. Since $(I-U)$ and $U$ have the same eigenvectors,  $\bf{v_{k}}$ is also the eigenvector of $U$ associated with the eigenvalue $\lambda_1$. 

The convergence of this classical algorithm is related to ratio of the two largest eigenvalues: i.e. in each iteration, the residual error in the estimated eigenvector decreases by a factor of $\frac{|1-\lambda_2|}{|1-\lambda_1|}$. 
In addition, the total required number of iterations depends on this ratio and scales as:
\begin{equation}
\label{EqIteration}
	O\left(\frac{n}{\log\frac{|1-\lambda_1|}{|1-\lambda_2|}}\right).
\end{equation}
This indicates that if the eigengap is  a polynomial of $n^{-1}$, and $||U||_2=O( poly(n))$, then the required number of iterations is  $O(poly(n))$.
In particular for an eigengap $2^{-m}$ and $||U||_2\leq 1$, the algorithm requires $O(n2^m)$ number of iterations.
\subsection{Quantum version}
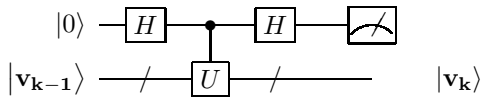
\begin{figure}
	\begin{center}
		\mbox{
			\Qcircuit @C=1em @R=.7em {
				\lstick{\ket{0}} & \gate{H} &  \ctrl{1} & \gate{H}&\qw & \meter \\
				\lstick{\ket{\bf{v_{k-1}}}} & {/} \qw &\gate{U}  &{/} \qw & \qw & \qw &\rstick{\ket{\bf{v_{k}}}}
				\\ 
			}
		}
	\end{center}
	\caption{\label{FigPowerMethod}Quantum circuit for the $k$th iteration of the shifted power method.}
\end{figure}

In the quantum version of this algorithm, we will first assume that $U$ is a unitary matrix with the eigenvalues $\lambda_j = e^{i\phi_j}$ and  for simplicity $\phi_j \in [0, \pi/2]$.  
For  the eigen-phases $\phi_1 \geq \dots \geq \phi_{2^n}$, this assumption maintains the same order for the magnitudes of the eigenvalues of $(I-U)$: i.e., $|1-\lambda_1| \geq \dots \geq |1-\lambda_{2^n}|$.  Therefore, the power iteration converges to the first eigenvalue.
The $k$th iteration of the classical method can be emulated by the quantum circuit presented in Fig.\ref{FigPowerMethod}.
As shown in the figure, the $k$th iteration in the algorithm starts with the initial state that involves the output vector of the previous iteration. The final quantum state before the measurement operation on the first qubit is as follows: 
\begin{equation}
\frac{1}{2}
\left(
\ket{0}\left(I+U\right)\ket{\bf{v_{k-1}}}+
\ket{1}\left(I-U\right)\ket{\bf{v_{k-1}}}
\right).
\end{equation}
After the measurement, the state on the second register collapses onto either $\frac{(I+U)\ket{\bf{v_{k-1}}}}{\alpha_{0k}}$ or $\frac{(I-U)\ket{\bf{v_{k-1}}}}{\alpha_{1k}}$ with probabilities respectively $\alpha_{0k}^2$ and $\alpha_{1k}^2$ which are the squares of the norms of the numerators. 
If every iteration, we collapse the state onto the same side (let us say the state where \ket{1} on the first register), then we basically implement the shifted power iteration described above. 
At the end, $\alpha_{1k}$ converges to $|1-\lambda_1|$ and $\ket{\bf{v_{k}}}$ is to the eigenvector of $U$. 
In addition, $\phi_1$ can be computed from the statistics of $\alpha_{1k}$ by using the cosine and sine components of $\lambda_1= \cos\phi_1+i\sin\phi_1$: 
Since $
(1-\cos\phi_1)^2 +  \sin^2\phi = \alpha_{1k}^2$ and $\cos^2\phi_1 +  \sin^2\phi = 1$,
\begin{equation}
\phi_1 = \arccos\left(1-\frac{\alpha_{1k}^2}{2}\right).
\end{equation}

Here, note that we have used $(I-U)$, however this can be generalized into $(\eta I -U)$ with a real coefficient $\eta$ by simply not using equal superposition state on the first qubit: i.e. using a different set of gates in lieu of the Hadamard gates on the first qubit.
This shifts the origin of the eigenvalues without changing the eigenvector and affects the convergence rate.
After computing one eigenpair, shifts along with other deflation techniques\cite{saad2011numerical} can be adjusted for the quantum power method as well to compute the rest of the eigenpairs.
Also note that when the dominant eigenvalue is multiplicative, the method will generate a superposition of the eigenvectors which can be used for principal component analysis. 

\subsection{Number of qubits}
Although the first register in Fig.\ref{FigPowerMethod} is drawn with 1 qubit, in the actual implementation the number of qubits can be as much as the parameter $maxIterate$. 
In addition, the convergence of the method can be determined by forming a simple state tomography of the measured qubit after every a few iterations and comparing it with the previous tomography. 
If it remains the same, then the iteration is stopped.
\subsection{Numerical experiment on the number of iterations}
\begin{figure}
\includegraphics[width=3in]{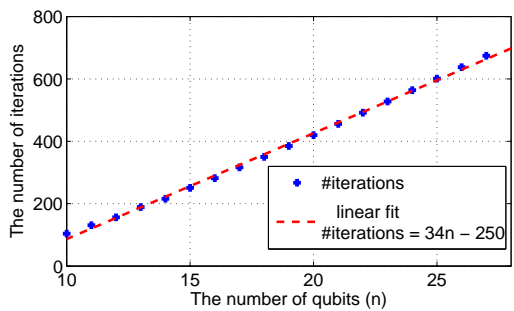}
\caption{\label{FigNIteration} For different $n$ values, the mean number of iterations for 15 runs for a fixed eigengap 0.01. The required number of iterations in general scales as in Eq.\eqref{EqIteration}.}
\end{figure}

\begin{figure}
\includegraphics[width=3in]{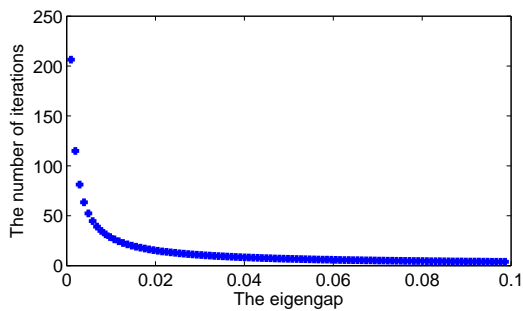}
\caption{\label{FigEigengap} For $n=20$ qubits, the number of iterations versus the eigengap.}
\end{figure}

As indicated before, the number of iterations is directly related to the eigengap. It is polynomial if the eigengap is polynomially small.
As a numerical example, we use a diagonal operator with random phases in $[0, \pi/3]$. Then we fix the eigengap (the difference between the first and the second phases) to 0.01. 
Then using the equal superposition state, we also set the initial probabilities of each eigenvector to $\frac{1}{2^n}$ which is exponentially small in the number of qubits.
Fig.\ref{FigNIteration} shows the mean number of iterations of 15 runs for each $n$ value in order to reach the success probability $\geq 0.5$ (the probability of having the dominant eigenvector on the second register). 
As seen in the figure, the number of iterations grows linearly with the number of qubits because of the fixed polynomial eigengap.
In Fig.\ref{FigEigengap}, we also show how the number of iterations is affected by the eigengap for $n=20$ qubits. 
As expected from Eq.\eqref{EqIteration}, the number of iterations to reach the success probability $\geq 0.5$ grows exponentially with the eigengap.

\section{Possible Applications}

\subsection{Eigenpair estimation}
On quantum computers, the phase estimation algorithm is the standard procedure for estimating an eigenpair of a given unitary matrix. The algorithm requires a good estimate for the initial state and uses the powers of the given matrix. 
Therefore, the quantum power method can be considered as an alternative to the phase estimation particularly when an initial good estimate for the eigenvector is not available or the powers of the matrix cannot be computed in an efficient way. 
However, when the gap between dominant eigenvalues is too small, the algorithm may require too many iterations. 

\subsection{Unconstrained Optimization} 
For $\bf{x}\in R^n$ and $f:R^n\rightarrow R$, $\min_{\bf{x}}f(\bf{x})$ is an unconstrained optimization problem that arises in the formulation of many problems. When $f$ is smooth, Newton based iterative methods are used in the solution of this optimization by starting  with an initial guess $\bf{x_0}$ and updating it toward the solution by using the gradient.

In the discrete optimization, $\bf{x}$ takes values from a discrete set such as $\{0,1\}^n$.
In terms of quantum states, we can simply formulate this as:
\begin{equation}
\label{Eqfindmin}
\min_{\ket{\bf{x}}}\ket{f(\bf{x})}.
\end{equation}

Representing each parameter $x_i$ with a qubit $q_i$, the solution space (the feasible set) can be described by the standard basis set $\{\ket{\bf{0}},\dots, \ket{\bf{2^n-1}}\}$.  Thus, $\{\ket{f(\bf{0})}, \dots, \ket{f(\bf{2^n-1})}\}$ describes the set of the associated fitness values for each basis (combination).
Suppose we have a mechanism $U_f$ to generate the following state:
\begin{equation}
\label{EqSolutionSpace}
\sum_{\bf{x}=\bf{0}}^{\bf{2^n-1}}\frac{1}{\sqrt{2^n}}\ket{f(\bf{x})}\ket{\bf{x}}.
\end{equation}
Then, after finding  the minimum $\ket{f(\bf{x})}$ with the Grover's search algorithm \cite{grover1997quantum,durr1996quantum}, we can obtain the corresponding combination for $x_i$s from the second register.
Grover's search in total requires $\approx \left( \frac{\pi\sqrt{N}}{4}\right)$ iterations for obtaining the solution with probability $\approx$ 1. 
Since any classical minimum finding algorithm would require $N$ operations in the worst case, there is a quadratic speedup in the quantum version. 
However, this speed-up is not sufficient to overcome the curse of dimensionality that inhibits finding exact solutions to many difficult problems. 

In our case, If we have a unitary operator $U_f$ with the eigenvalues $\exp(if(\bf{x}))$  and associated eigenvectors \ket{\bf{x}}s, then the solution of Eq.\eqref{Eqfindmin} becomes equivalent to finding the groundstate of the Hamiltonian $H$ in $U=e^{iH}$.
Or if we are given $U_f$ that generates Eq.\eqref{EqSolutionSpace},   using the quantum Fourier transform,  $\ket{f(\bf{x})}$ can be mapped to the phases $\exp(if(\bf{x}))$. 
Then the quantum power method can be used to find maximum $f(\bf{x})$ and its associated eigenvector $\ket{\bf{x}}$. This may provide exponential speed-up over the search algorithm when the eigengap is $\Omega(poly(1/n))$. 

 \subsection{Quadratic unconstrained binary optimization }
A quantum computer cannot solve a general instance of an NP problem in polynomial time unless polynomial-hierarchy in the complexity theory collapses completely. 
However, it may allow more efficient optimization and approximation algorithms over the known classical ones.
Many NP problems can be formulated as a quadratic unconstrained optimization problem (QUBO) used in adiabatic quantum computers  (e.g.\cite{lucas2014ising}). 
Here, without describing a particular NP problem, we will show how to map the general formulation of QUBO into eigenvalues of a diagonal matrix $U_f$. As in the maximum finding, the quantum power method can be used to extract a maximal solution. Note that this requires exponentially large number of steps when the eigengap is the exponential of $n$, the number of parameters.

QUBO is defined as maximizing or minimizing the following equation:
\begin{equation}
H = \sum_{j=0}^{n-1}c_jx_j+ \sum_{j=0}^{n-1}\sum_{k=j+1}^{n-2}q_{jk}x_jx_k
\end{equation}
where $c_j$ and $q_{jk}$ are given real coefficients and $x_j$ and $x_k \in \{0,+1\}$.  
Considering $x_j$ and $x_k$ as $\in\{-1,+1\}$, this problem is mapped to the Ising type Hamiltonian used in the adiabatic quantum computation \cite{farhi2000quantum,farhi2001quantum}.

For mapping to a diagonal unitary, we represent each parameter by a qubit. Then, we use one qubit gates $Z_j$s for the summands in the first summation and two qubit gates $Z_{jk}$s for the ones in the second summation:
We define a one-qubit gate for the summand in the first summation as:
\begin{equation}
Z_j = \left(
\begin{matrix}
e^{ic_j \alpha\pi} & 0\\
0 & e^{ic_j \beta\pi}
\end{matrix}
\right), 
\end{equation}
where either $\alpha = 0$ and $\beta = 1$
or $\alpha = 1$ and $\beta = -1$. 
Then two qubit controlled gate is defined for the terms in the second summation:
\begin{equation}
Z_{jk} = \left(
\begin{matrix}
1&0&0&0\\
0&1&0&0\\
0&0&e^{iq_{jk} \alpha\pi} & 0\\
0&0&0 & e^{iq_{jk} \beta\pi}
\end{matrix}
\right).
\end{equation}

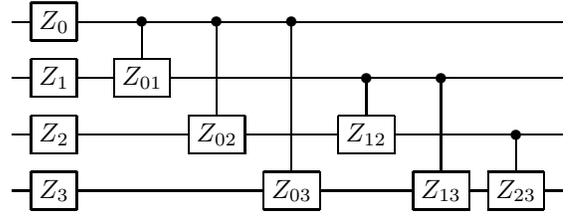
\begin{figure}
	\begin{center}
		\mbox{
			\Qcircuit @C=0.7em @R=.7em {
				&	 \gate{Z_0} 	&	\qw	&	\ctrl{1}	&	\ctrl{2}	& \ctrl{3}	&	\qw	&	\qw	&	\qw	&	\qw	\\	
				&	 \gate{Z_1} 	&	\qw	&	\gate{Z_{01}}	&	\qw	&	\qw	&	\ctrl{1}	&	\ctrl{2}	&	\qw	&	\qw	\\
				&	 \gate{Z_2} 	&	\qw	&	\qw	&	\gate{Z_{02}}	&	\qw	&	\gate{Z_{12}}	&	\qw	&	\ctrl{1}	&	\qw	\\
				&	 \gate{Z_3} 	&	\qw	&	\qw	&	\qw	&	\gate{Z_{03}}	&	\qw	&	\gate{Z_{13}}	&	\gate{Z_{23}}	&	\qw	\\
			}
		}
	\end{center}
	\caption{\label{FigQUBO}The explicit circuit for the unitary operator of QUBO illustrated for $n=4$ qubits: For $0 \leq j,k \leq n-1$; while $Z_{j}$ represents a phase gate on the $j$th qubit, $Z_{jk}$ is a standard two qubit controlled phase gate on the $j$th and $k$th qubits. There are $\frac{n^2+n}{2}$ quantum gates in total. 
	}
\end{figure}
The circuit in Fig.\ref{FigQUBO} represents the alignment of these quantum gates for 4 parameters. 
This circuit is equivalent to a diagonal matrix 
whose eigenvectors indicate the possible candidate solutions and eigenphases are the fitness values of these candidates. 
Using an initial superposition state, the quantum power method can be used to extract the maximum (or minimum) of this problem in $O(poly(n))$ time if the eigengap is $\Omega(poly(1/n))$.

\subsection{Finding groundstates of Hamiltonians}
A k-local quantum Hamiltonian can be considered as a binary optimization by mapping Pauli spin matrix $\sigma_z$ into  their eigenvalues $\{-1,1\}$. Hence, using the QUBO formulation, one can also try to find the groundstate of the Hamiltonian in the following form:
\begin{equation}
H = \sum_{i,j>i}^{n}J_{ij}\sigma_z^i \sigma_z^j,
\end{equation}
where $J_{ij}$s are coefficients and $\sigma_z^i$ indicates a Pauli spin operator on the $i$th qubit.
\subsection{ Generalization to the quadratic assignment problem}
 Quadratic assignment problem is another formulation used to represent many discrete NP-hard optimization problems such as traveling salesman problem.
 Given three matrices $F = [f_{ij}]$, $D=[d_{kp}]$, and $B = [b_{ik}]$ representing respectively the flow between facilities, the distances between locations, and the allocation costs to the locations;  the general form of the quadratic assignment problem of order $n$ is  defined as follows\cite{loiola2007survey,pardalos1994quadratic}:
 \begin{equation}
 \min	\sum_{i,j = 1}^n \sum_{k,p = 1}^n f_{ij}d_{kp}x_{ik}x_{jp} + \sum_{k,p = 1}^nb_{ik}x_{ik},
 \end{equation} 
 such that:
 \begin{equation}
 x_{ij} \in \{0,1\} \text{\ for\ } 1 \leq i, j\leq n,
 \end{equation}
 \begin{equation}
 \sum_{i = 1}^n x_{ij}=1 \text{\ for\ } 1 \leq j\leq n, 
 \end{equation}
 and 
 \begin{equation}
 \sum_{j = 1}^n x_{ij}=1 \text{\ for\ } 1 \leq i\leq n. 
 \end{equation}
 In this formulation, $(f_{ij}d_{kp})$ indicates the cost to simultaneously assign the facilities $i$ and $j$ to the locations $k$ and $p$. Here, note that disregarding the linear term above does not simplify the problem \cite{loiola2007survey}.
 
 This problem has also many different mathematical formulations such as trace and Kronecker product: 
 The trace formulation is defined by the following objective function:
 \begin{equation}
 \min_{X\in \Pi} f(X) = trace(FXD+BX^T),
 \end{equation}
 where $\Pi$ is the set of permutation matrices and $X^T$ denotes the transpose of the matrix $X$.
 Considering the vector form, $vec(X)$ of the matrix $X$, the objective function can be  rewritten in the following equivalent Kronecker product form:
 \begin{equation}
 \min vec(X)^T\left(F\otimes D\right)vec(X) + vec(B)^Tvec(X)
 \end{equation}

\subsubsection{Formulation of QAP with a quantum circuit}
 To get the solution from eigenstate, we need to use $n^2$ number of qubits each of which represents an assignment. In the output if the qubit  $x_{ij}$ is 1, then  facility $i$ is assigned to the location $j$. Here, we basically have the same formulation as in QUBO.
 However, in this case, we use $n^2$ qubits and $\frac{n^2(n^2-1)}{2}+n^2 = \frac{n^4+n^2}{2}$ number of operations in total.
 
\section{Conclusion}
In this paper, we have sketched the steps for the quantum power method and discussed its possible applications. 
In addition to any eigenvalue related problems, the method can be used to find maximum/minimum of a quantum state in polynomial time if the eigengap is sufficiently large. 
With the polynomial mapping, it can be also used as a part of an optimization algorithm to solve many difficult problems. 
In particular, we formulate the quadratic unconstrained binary optimization and quadratic assignment problem as finding the minimum eigenvalue of a diagonal operator.
While on classical computer this formulation requires $O(2^n)$ space, it can be simulated on quantum computers by using $O(poly(n))$ qubits and operations. 
Although the quantum version can be implemented efficiently on a possible quantum computer based on open quantum systems, the required number of qubits growing linearly with the number of iterations would inhibit the simulation of the method on the current quantum computer technologies for nontrivial $n$ values.


\end{document}